\documentclass{aastex}
\usepackage{graphicx}
\usepackage{emulateapj5}
\usepackage{onecolfloat}

\newcommand{\um}{\mbox{{\usefont{U}{eur}{m}{n}\symbol{22}}m}}

\lefthead{Kranz, Slyz \& Rix}
\righthead{Probing for Dark Matter in Galaxy Disks}
\slugcomment{Accepted to the Astrophysical Journal}

\begin{document}

\twocolumn[

\title{Probing for Dark Matter within Spiral Galaxy Disks}

\author{Thilo Kranz, Adrianne Slyz and Hans-Walter Rix}

\affil{Max-Planck-Institut f\"ur Astronomie, K\"onigstuhl 17, 69117 Heidelberg, Germany}

\email{kranz@mpia.de, slyz@mpia.de, rix@mpia.de}

\begin{abstract}
We explore the relative importance of the stellar mass density as compared to
the inner dark halo, using the observed gas kinematics throughout the disk of
the spiral galaxy NGC 4254 (Messier 99). We perform hydrodynamical simulations of the
gas flow for a sequence of gravitational potentials in which we vary the stellar disk
contribution to the total potential. This stellar portion of the potential was derived
empirically from color corrected K-band photometry reflecting the spiral arms in the
stellar mass, while the halo was modelled as an isothermal sphere. The simulated gas
density and the gas velocity field are then compared to the observed stellar spiral arm
morphology and to the H$\alpha$ gas kinematics. We find that this method is a powerful
tool to determine the corotation radius of the spiral pattern and that it can be used to
place an upper limit on the mass of the stellar disk. For the case of the galaxy NGC 4254
we find ${\rm R}_{\rm CR} = 7.5 \pm 1.1\, {\rm kpc}$, or
${\rm R}_{\rm CR} = 2.1\,{\rm R}_{\rm exp}$(K'). We also demonstrate that for a maximal
disk the prominent spiral arms of the stellar component over-predict the non-circular gas
motions unless an axisymmetric dark halo component contributes significantly ($\ga 1/3$)
to the total potential inside 2.2 K-band exponential disk scale lengths.
\end{abstract}
\keywords{galaxies: kinematics and dynamics --- galaxies: structure --- galaxies: halos
--- galaxies: spiral --- galaxies: individual (NGC 4254)}

 ]

\section{INTRODUCTION} 
In almost all galaxy formation scenarios non-baryonic dark matter plays
an important role. Today's numerical simulations of cosmological structure
evolution reproduce fairly well the observed distribution of galaxy properties
in the universe \citep{kau99} and attempts to model the
formation of single galaxies have been made as well \citep{ste95}. 
In these simulations the baryonic matter cools and settles in the
center of dark halos where it forms stars. The distribution of stars and gas
in a galaxy depends strongly on the local star formation and merging history. 
At the same time that the stars are forming the halos evolve and merge as well.

The final relative distribution of
luminous and dark matter in the centers of the resulting galaxies is under debate
because the mass distribution of the dark matter component is difficult to assess directly.
Measuring luminous and dark matter mass profiles separately requires
innovative strategies because the halo is poorly constrained and equally good fits to
measured rotation curves can be achieved for a wide range of
visible mass components \citep{bro97}. In order to define a
unique solution to this so called 'disk-halo degeneracy', the 'maximal disk'
solution was introduced. It assumes the highest possible mass-to-light ratio (M/L) for
the stellar disk \citep{val85,val86}.
A practical definition is given by \citet{sac97} who attributes
the term 'maximal' to a stellar disk if it accounts for 85\% $\pm$ 10\% of the total
rotational support of the galaxy at R = 2.2 R$_{\rm exp}$. This approach
has proven to be very successful in matching observed \ion{H}{1} and H$\alpha$ rotation curves
\citep{val85,ken86,bro97,sal99} and also satisfies some dynamical constraints, such as the
criteria of forming $m = 2$ spirals \citep{ath87} as well as 
observational constraints on the structure of the Milky Way \citep{sac97}.
However, modern numerical 
N-body simulations find significant central dark matter density cusps \citep{fuk97,moo99}.
Even if the prediction of these strong density cusps
may not be entirely correct, the simulations find that the dark matter is of comparable importance
in the inner parts of galaxies \citep{blu86,moo94,NFW1,NFW2} and it thus has a considerable
influence on the kinematics. In this case a stellar disk of a galaxy would turn out
to be 'sub-maximal'. 

\begin{figure}[t]
 \resizebox{\hsize}{!}{\includegraphics{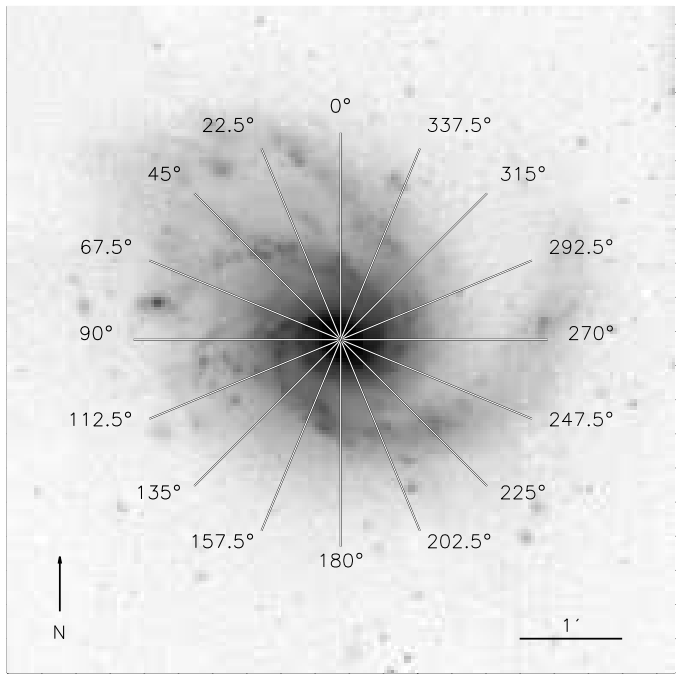}}
 \caption{K'-band Image of NGC 4254 with a total exposure time of 20 minutes at the Calar Alto
3.5m telescope. Bright foreground stars are masked out. The overlay shows the slit orientations
of the spectrograph. We took 8 longslit spectra (angles labelled in bold font) crossing the
galaxy's center to measure the 2D velocity field.
 \label{n4254strategy}}
\end{figure}

It is important to determine the relative proportion of dark and luminous matter
in galaxies for a better understanding of the importance of the baryonic mass
in the universe. This proportion also bears information on the dynamics and structure
of the dark matter itself. Spiral galaxies are well suited to study dark matter distributions because
their distinctly ordered kinematics provide an excellent tracer of the gravitational
potential in the disk plane. Since bars in galaxies are very prominent features with
distinct dynamic characteristics, they are especially well suited to evaluate the amount of
luminous matter. Sophisticated studies of barred galaxies indicate that their stellar disks
alone dominate the kinematics of the inner regions -- the stellar contribution is 'maximal' 
\citep{deb98,deb00,wei01b}.
However, studies of our own Milky Way, maybe also a barred spiral, still do not give a clear answer
as to whether the disk is maximal \citep{sac97,eng99} or not \citep{kui95,deh98}.
Bottema's analysis of the stellar velocity dispersion in various galactic disks
led to the conclusion that disks cannot comprise most of the mass inside the radial
range of a few exponential scale lengths \citep{bot97}. 
Aside from the dynamical analysis of single systems, other attempts to tackle this problem
have been undertaken. \citet{mal00} used the geometry
of gravitational lens systems to probe the potential of a lensing galaxy.
They concluded that a maximum disk solution is highly unlikely.
\citet{CR99} applied statistical methods to learn 
about the mass distribution in galaxies. In their analysis
they found no dependence of the maximum rotation velocity on a galaxy's disk size. This
is considered to be a strong argument to rule out a maximum disk solution. The conflicting
findings of different studies leave the question of the relative proportion of dark and
luminous matter in galaxies still open.

In this paper we want to exploit the fact that the stellar mass in
disk galaxies is often organized in spiral arms, thus in kinematically cold
non-axisymmetric structures. In the canonical CDM cosmology the dark matter
is collision-less and dominated by random motions. Although the introduction of
weakly self-interacting dark matter was proposed to avoid current shortcomings of
the CDM model \citep{spe00} it seems to raise other, comparably severe problems
\citep{mir00,ost00}. Hence it seems reasonable to assume that CDM is not substantially 
self-interacting, but dynamically hot and thus not susceptible to spiral and
non-axisymmetric structure. 

In light of this, the key to measuring the baryonic to dark matter fraction is to make use of the
non-axisymmetric structure that we can observe in the stellar light distribution.
 Using deviations from axisymmetry of stellar disks, several efforts have
already been made to constrain the dark matter content of galaxies (eg., Visser 1980; Quillen 1999, and references
therein). Some of the most significant conclusions came from studies of massive bars, which are
the strongest non-axisymmetric structures in disk galaxies. Spiral arms comprise a less
prominent, but still significant mass concentration. Already very early theoretical
calculations of gas shocking in the gravitational potential of a spiral galaxy \citep{rob69}
told us that we could expect ``velocity wiggles'' with
an amplitude of 10 to $30\, {\rm km\,s}^{-1}$ while crossing massive spiral arms. For ionized gas,
measurements of the velocity to this precision can be achieved with common
longslit spectrographs. The imprint of the spiral pattern
in the velocity field of real galaxies is indeed not very strong, as apparent in the 2D velocity fields
of M100 \citep{can97}, of low surface brightness galaxies \citep{qui97} or of a sample of spiral
galaxies \citet{sak99}. There are only a few spiral galaxies without bars that show stronger
wiggles in the velocity field that are associated with the arms, eg. M81 \citep{vis80,adl96} and M51
\citep{aal99}. In order to still achieve the goal of measuring mass-to-light ratios we need
to compare the weak features that we expect in the measured velocity field to detailed kinematic models.
Using new high resolution K-band photometry to map the stellar component
and employing a modern hydro-code to simulate galactic gas flows we are confident that our
models show enough details and we can eventually measure mass-to-light ratios.
If the arms are a negligible mass concentration
relative to the dark matter distribution in the galaxy, these wiggles should appear only very barely
in the velocity field. The main aim of this project is to find out what fraction of
the rotation speed comes from a mass component with spiral arms. In order to do this
we compare the strength of the wiggles in a galaxy's
observed velocity field to a model of the gas velocity field arising in a potential
whose disk-halo fraction is known. As input for our gas dynamical simulations, we derive
the stellar potential of the galaxy from color-corrected K-band photometry, while the dark matter component
is modelled as an isothermal sphere with a core. Simulations are performed for a variety of
potential combinations and values for the pattern speed of the spiral structure. The results
from these simulations are then compared to the observed kinematics.

The paper is organized as follows. In the next Sections we present the galaxy sample and the
observations (\S2) and the general modelling techniques (\S3). In Section 4 we present
results for the galaxy NGC 4254 (Messier 99) and discuss them in \S5. We summarize in \S6.

\section{THE SAMPLE DATA AND NGC 4254}
We aim to perform this study on a sample of about half a dozen galaxies to 
account for systematic errors in the analysis of single systems.
We could not identify a data set in the literature that fulfilled satisfactorily the
requirements for this project. Hence we decided to collect data on a sample of galaxies ourselves.
The data acquisition procedure
was carried out in the following way. First we took NIR photometry of possible
sample candidates to check whether their spiral structure shows up reasonably well 
in the K-band. From that sample we selected the most promising candidates and 
obtained kinematic data for them on a second observing run.
 
The requirement for the kinematic measurements was to trace the gas velocity perturbations
of the spiral arms, ideally across the whole disk. The two classical methods for obtaining
2D gas velocity fields are \ion{H}{1} or CO radio observations and Fabry-Perot interferometry.
Alternatively, longslit spectra taken at different position angles can be used to map the
disk. 

Single dish \ion{H}{1} or CO observations are not suited for our project because they suffer 
from relatively bad angular resolution and poor sensitivity to faint emission between spiral
arms and the outer part of the disk. \citet{sak99} recently published CO observations
of a sample of spiral galaxies. In the velocity maps the signature of the spiral arms is in
the majority of cases not or only barely visible. This most likely results from beam smearing
of the anyway weak velocity perturbations. 2D Fabry-Perot velocity fields provide the required
angular resolution but usually give only a very
patchy representation of the disk: Mainly the \ion{H}{2} regions show up in the map
\citep{can93,reg96,wei01a}. Because the coverage we can achieve by taking eight longslit
spectra across a galaxy's disk is reasonably high, in combination with its good angular
resolution and sensitivity to faint H$\alpha$-emission we chose to collect our kinematic
information by taking longslit spectra.

This paper deals only with the results of the first galaxy from the sample: NGC 4254.

\subsection{Observations and Data Reduction}
Since we require photometric as well as kinematic data for this study, there are 
some constraints which apply for the sample selection. To reasonably resolve the structures
of the stellar disk we prefer galaxies with low inclinations with respect to the line-of-sight (LOS).
However, the LOS component of the circular motion increases with
inclination \emph{i}. Due to the fact that the projection of the galaxy scales with cos(\emph{i})
and the LOS fraction of the velocity scales with sin(\emph{i}), 
galaxies in the inclination range between 30\degr ~and 60\degr ~are best suited for yielding all
the photometric and spectroscopic information. We also chose the sample to consist mostly of
non-barred high luminosity galaxies with strong spiral arms.
All data were acquired at the Calar Alto observatory's 3.5\,m telescope.

For all sample galaxies we decided to take near-infrared (NIR) photometry to study the
luminous mass distribution and gas emission line (H$\alpha$) spectroscopy to acquire kinematic
information for the systems.
The NIR images were taken during two observing runs in May 1999 and March 2000 with the Omega
Prime camera at the 3.5\,m's prime focus \citep{biz98}. It provides a field
of view of 6\farcm76 $\times$ 6\farcm76 with a resolution of 0\farcs3961 per pixel. We used
the K' filter which has a central wavelength of 2.12 \um. The integration time was 20 minutes
on target which allows us to trace disks out to about 4 scale lengths. Our observation 
sequence included equal amounts of time on the targets and sky frames used for background
subtraction. The data reduction was performed using standard procedures of the ESO-Midas
data reduction package. In total, we collected NIR photometry data for 20 nearby spiral
NGC-galaxies. From their NIR appearance we selected half of the galaxies as a subsample for
which kinematic data should be taken.

We obtained the gas kinematics from longslit spectroscopy measurements of the H$\alpha$ emission.
With the setup we used, the TWIN spectrograph achieved a spectral resolution of 0.54\,\AA~per
detector pixel, which translates to $24.8\,{\rm km\,s}^{-1}$ LOS-velocity resolution per pixel,
allowing us to determine LOS-velocities with $\sim 7\,{\rm km\,s}^{-1}$ precision. We sampled the
velocity field of the galaxy along 8 slit
position angles\footnote{All position angles quoted in the text are in degrees eastward from north},
all crossing the center of the galaxy (cf. Fig.~\ref{n4254strategy}). The
slit of the TWIN spectrograph measures 4\arcmin $\times$ 1\farcs5 on the sky. The spectra span
a region of 6130 -- 7170\,\AA~centered on H$\alpha$. The spectra were obtained during three
observing runs in June 1999, May and December 2000. The integration time for each single
spectrum varied between 600\,s and 1800\,s due to weather and scheduling constraints.
For the data reduction we used the IRAF package and applied standard longslit reduction
procedures. We determined the LOS velocity component of the ionized gas as a function
of radius from the galactic center
from doppler shifts of the H$\alpha$-line. The center of the emission line was 
determined by fitting separately a single Gaussian profile to it and the brighter \ion{N}{2}-line at 6584\,\AA.
The weighted comparison of the two fits provided us with the uncertainty of the line center position.
Finally, each of the eight spectra were folded at the center of the 
galaxy to get the two rotation curves. To do this coherently, we determined an
average wavelength distance $D_{\lambda}$ between a prominent sky line (at $\lambda_{sky}$) and H$\alpha$
at our best guess for the galactic center from all eight spectra. Then, separately for each slit,
the H$\alpha$ rest wavelength in the center of the galaxy -- where $v_c = 0$ -- was assigned
to be at $\lambda_{sky} + D_{\lambda}$. There the two rotation curves should be separated.
This provided us finally with 16 rotation curves at different position angles per
galaxy, each reaching out to about 2 arc minutes. All the 16 rotation curves are displayed later in Section~4,
Fig.~\ref{allslits} together with the results from one simulation.

\subsection{Structural Properties of NGC 4254}\label{morphology}
We selected NGC 4254 (Messier 99) as the first galaxy from our sample to be analyzed,
because it shows a clear spiral
structure with high arm-inter-arm contrast. NGC 4254 is a bright Sc I galaxy located in
the Virgo galaxy cluster with a recession velocity of $2407\,{\rm km\,s}^{-1}$ adopted from
NED\footnote{The NASA/IPAC Extragalactic Database (NED) is operated by the Jet Propulsion
Laboratory, California Institute of Technology, under contract with the National Aeronautics
and Space Administration.}. 
We assume a distance of 20 Mpc towards NGC 4254, taken from the literature \citep{san76,pie88,fed98}.
It has a total blue magnitude of $B_T = 10.44$ and a diameter of 5.4
$\times$ 4.7 arcmin on the sky. At 20 Mpc one arc second is 97 parsecs in the galaxy
which translates to 38.4 pc per detector pixel. Our H$\alpha$ rotation curve for NGC 4254
(the H$\alpha$ rotation curve is plotted later in Fig.~\ref{rc_combin}
together with modelled rotation curves) rises steeply out to $\sim 35$\arcsec\ (3.4 kpc) and then
flattens at a rotation velocity of $\sim 155\,{\rm km\, s}^{-1}$. This agrees well with
earlier estimates \citep{pho93}. From the kinematics we know that the
south-western part of the galaxy is approaching. If we assume trailing spiral arms then
the galaxy rotates clockwise when viewed from our perspective.

\begin{figure*}[t]
 \resizebox{\hsize}{!}{\includegraphics{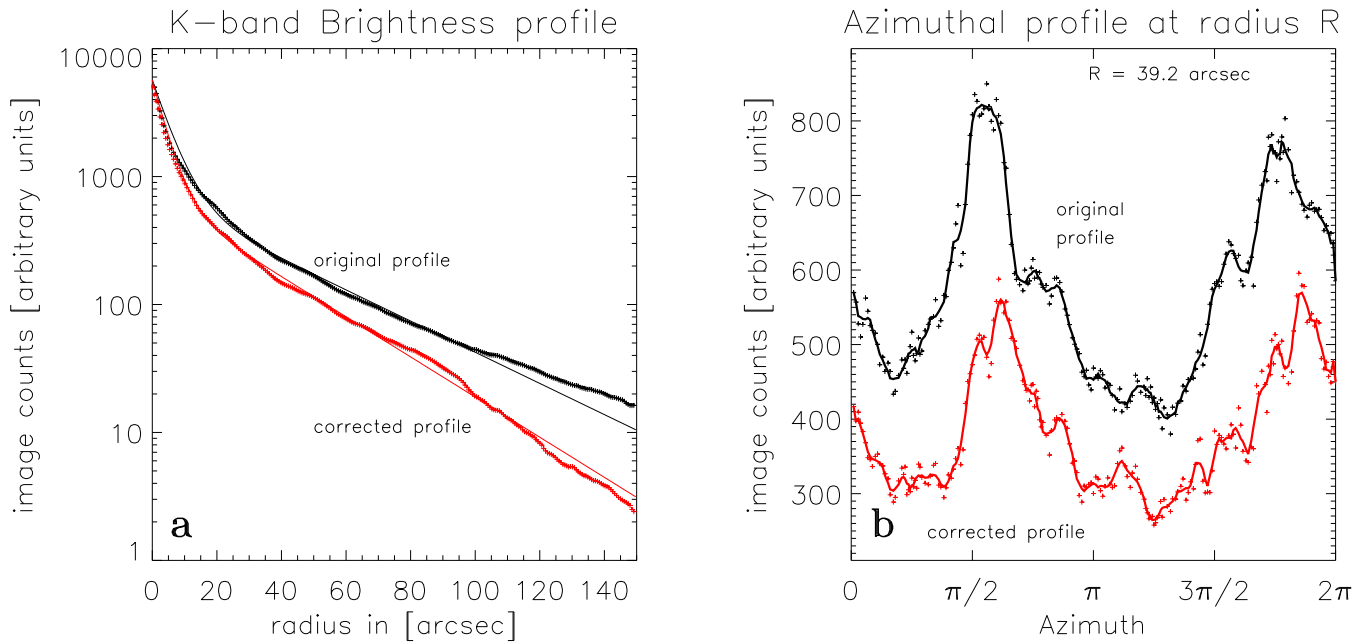}}
 \caption{Different results from the point-by-point correction of the mass-to-light ratio
in the K' image. (a) The original azimuthally averaged brightness profile together
with the corrected profile. The outward blueing of the galaxy leads to a stellar disk scale
length which is smaller than in the uncorrected K'-band light. (b) The azimuthal brightness
profile at a radius of 39\farcs2; azimuth 0 is to the north. The profile shows a clear two
arm structure that is preserved after the correction. Note, however, that the shape of the
profile changes.
 \label{mlcres}}
\end{figure*}

In the K-band NGC 4254 shows very prominent two-arm spiral features at most radii. The
northern arm bifurcates at R $\approx 4.5$ kpc, causing a three arm pattern in the
outer disk. Furthermore, the galaxy exhibits considerable lopsidedness.
NGC 4254 shows a strong arm-interarm
brightness contrast, noted by \citet{sch76} to be even stronger than the one for 
NGC 5194 (Messier 51). Even in the K-band the brightness contrast is rather high, approximately
a factor of 2 over a wide radial range. \citet{gon96} argued 
that, combined with a usual density wave, some external mechanism is needed to
invoke such high contrasts. However, NGC 4254 is well separated from any other galaxy in the
cluster and there are no obvious signs for recent interaction. \citet{pho93} reported
in a recent paper the detection of high velocity \ion{H}{1} clouds
outside the disk's \ion{H}{1} emission. The authors argue that in-falling \ion{H}{1} gas may be
responsible for the unusual 'one armed' outer structure of the spiral. In that case
the spiral structure of NGC 4254 may have been recently reorganized, enhancing the southern
arm or the arm-interarm mass density distribution.
Recent interactions could in principle corrupt the project's assumption of a steady state
spiral pattern. But as we will find from our full analysis, the steady state assumption
is seemingly not far off on time scales of a few dynamical periods.

NGC 4254 harbors a small bar-like structure
at its center with a major axis position angle of $\approx 40$\degr. From both
ends of the bar two major arms emerge with a third arm splitting off
the northern arm. By analyzing a $g-K$ color map of NGC 4254 one learns that this third 
arm is significantly bluer than the other regions of the galaxy and thus consists of
a younger stellar population.

In the K-band the disk of NGC 4254 is well approximated by an exponential with a scale 
length of R$_{\rm exp} \, \approx$ 36\arcsec, corresponding to $\approx$ 3.5 kpc if a
brightness average of the arm and interarm regions is
considered and bright \ion{H}{2} regions are removed from the image. Except for the very center,
the whole surface brightness profile is well fitted by a double exponential model with an
inner 'bulge' scale length of $\approx$ 0.6 kpc (cf.~Fig.~\ref{mlcres}a)

\section{MODELLING}

In the forthcoming Sections we describe the modelling needed to connect our photometric and
kinematic observations. In this pilot study, we restrict our analysis and description to
the spiral galaxy NGC 4254 (Messier 99).
The data modelling involves four discrete steps. First we derive a stellar gravitational
potential from the K-band images. Second, for each stellar mass-to-light ratio, we find
a dark matter halo profile to match the total rotation curve. Third, we perform
hydrodynamical simulations of the gas within the combined stellar
and dark matter potential. Finally, we compare the predicted gas velocity field to our
H$\alpha$-observations.

\subsection{The Stellar Potential} \label{modstarpot}
\subsubsection{K-band Light as Tracer of the Stellar Mass}\label{k_light}

\begin{figure*}[t]
 \resizebox{\hsize}{!}{\includegraphics{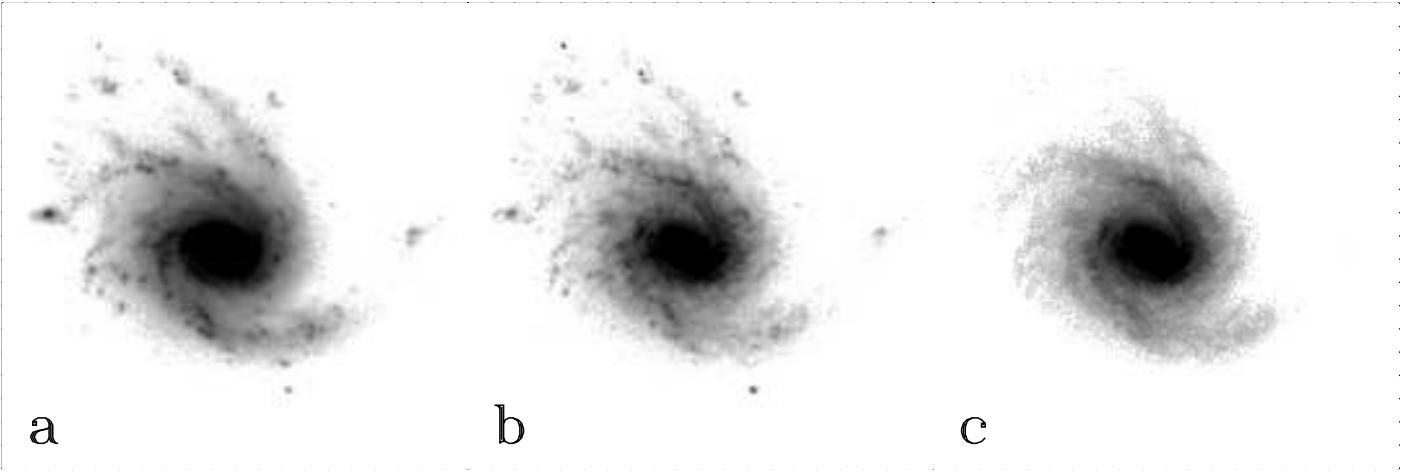}}
 \caption{Different stages of processing of the K-band image prior to the calculation 
of the potential. (a) The original frame. (b) Image of the galaxy after
the correction for constant mass-to-light ratio was applied. (c)
Final image, after removal of compact star forming regions by means of
azimuthal Fourier decomposition. The images are not yet deprojected to face-on.
 \label{mlcorr}}
\end{figure*}

We would like to derive the stellar potential directly from the deprojected K-band stellar
luminosity density of the galaxy. To this end,
it is important to understand how well the NIR surface brightness traces the stellar mass
density. There are two major concerns which complicate the direct translation: both dust extinction
and spatial population differences could introduce K-band mass-to-light ratio variations. 
Observing at $\sim 2\,$\um \, reduces the effect of the dust extinction significantly. Since we look
at most galaxies from a nearly face-on perspective, we expect the optical depths to be
$\sim 0.5$ in the K-band, leading to local flux attenuations of $< 10$\% \citep{RR93,rix95}.
Significant global M/L-variations may arise from red supergiants; they emit 
most of their light in the near infrared and are fairly numerous and thus could considerably
affect the total light
distribution in K-band images. Supergiants evolve very rapidly and therefore
are mainly found within the spiral arms where they were born. This would tend to 
increase the arm-interarm contrast, leading to stronger inferred spiral perturbations in the stellar
potential.

We used the approach of \citet{bel01} to
correct for local mass-to-light ratio differences, which may still be present in the K-band images.
From spiral galaxy evolution models, assuming a universal IMF, \citet{bel01} find
that stellar mass-to-light ratios correlate tightly with galaxy colors.
They provide color dependent correction factors, which we can use to adjust our K-band
image to constant mass-to-light ratio. For NGC 4254 we used a $g$-filter image taken from 
Z. Frei's galaxy catalog \citep{fre96}. 
The $g$-filter \citep{thu76} has a passband that is very similar to the Johnson
$V$-filter, so we can use the color correction given for $V-K$ while using $g-K$ colors.
The mass-to-light ratio is given by
\begin{equation}
\label{mlcformel}
\log_{10}(M/L) = -1.087 + 0.314 (V-K).
\end{equation}
This relative correction was performed on a pixel by pixel basis by scaling each flux by
the corresponding $M/L$ (cf. Fig.~\ref{mlcorr}b). NGC 4254 has 
V-K of 2.61 which results, according to (\ref{mlcformel}), in a K-band maximal-disk stellar M/L,
$\Upsilon_{\star}$, of 0.54. If we estimate a maximal-disk stellar M/L from the disk potential
we use for the simulations, we find $\Upsilon_{\star} \approx 0.64$. These values
are in good agreement, considering the uncertainties that contribute to the estimation of
$\Upsilon_{\star}$. In our approach even small changes in NGC 4254's distance or inclination
result in errors of $\approx 20\%$.  
The galaxy becomes significantly bluer from the center
to the outer disk regions, with the bluest regions being the spiral arms themselves. As
a result the mass density profile of the corrected disk gets steeper: the disk scale length
(after deprojection) changes from 3.5 kpc to 2.7 kpc (cf. Fig.~\ref{mlcres}a).
In general the correction for
constant mass-to-light ratio does not particularly change the arm-interarm contrast of NGC 4254
and thus does not reduce the non-axisymmetric component of the stellar potential. In the force
field of the galaxy, the relative strength of the tangential component makes up $\sim 6$\%
of the radial force component before and after the correction. Nevertheless, the correction
modifies the non-axisymmetric force component on average by $\sim 30$\% (cf. Fig.~\ref{mlcres}b).
This has an evident
effect on the data and the models, which we think is favorable.
The correction seems to work fine for most of the disk. However, there are some thin dust
lanes at the inner edges of the spiral arms in the green image of NGC 4254 that are
optically thick. Because of the quite poor resolution of the green image, the dust lanes are not
distinctly visible in the image. For these the correction leads to some overcorrection because
virtually all the green
light is absorbed by the dust lane. But since these optically thick dust lanes are 
narrow and still at the location of the NIR spiral arms, we assume that their influence
on the total potential is negligible.

To reduce the influence of star forming \ion{H}{2}-regions and OB-associations, which are 
distinctly visible in the K-band data, we filtered the image by means of a Fourier decomposition.
From the azimuthal decomposition we discarded the Fourier terms higher than N=8 and subtracted
the residual from the K-band image of NGC 4254. This correction
does not depend significantly on the number of Fourier terms included in the fit.
This procedure cleaned the image of the patchy small scale structure and left us with the
smooth global spiral pattern (cf. Fig.~\ref{mlcorr}c).

\subsubsection{Deprojection}

\begin{figure*}[t]
 \resizebox{\hsize}{!}{\includegraphics{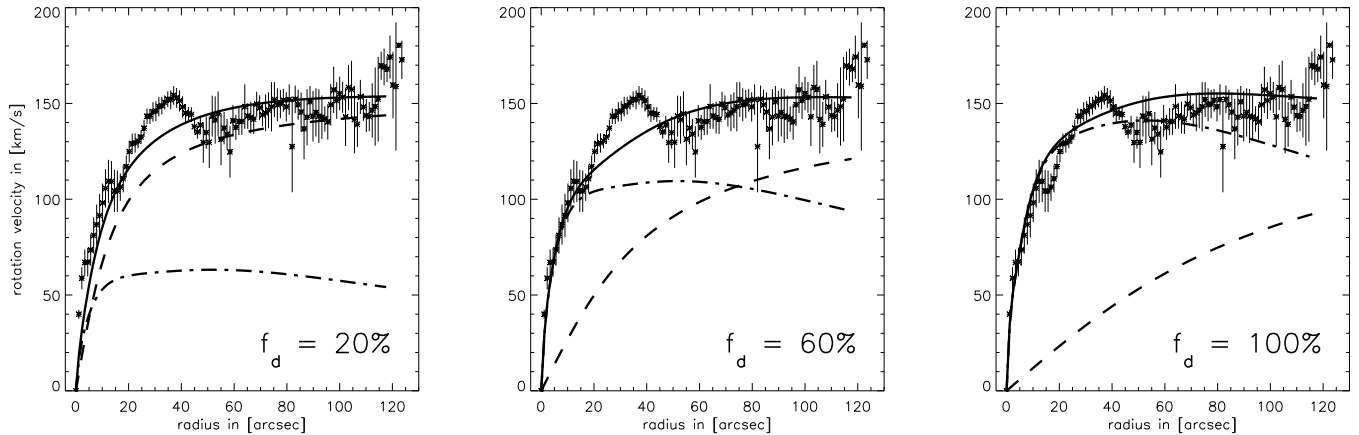}}
 \caption{Different decompositions of the total gravitational potential into stellar and dark halo
  components, illustrated for the example of NGC 4254. The measured rotation curve (data points with
  error bars) -- an average of six single rotation curves closest to the major axis -- is well fit
  by all decompositions (solid line {\bf ---}). Displayed are three cases for the contribution of the
  stellar disk f$_{\rm d}$ = 20\%, 60\% and 100\% (dash-dotted line {\bf -- $\cdot$ --}). The halo
  contribution (dashed line {\bf -- --}) was adjusted to yield a good fit to the measurements in
  combination with the pre-selected stellar contribution. The fit parameters for the three cases are
  listed in Table \ref{potpar}.
 \label{rc_combin}}
\end{figure*}

To deproject the K-band images to
face on, we need to derive the disk inclination, \emph{i}, and the major axis
position angle, PA from the observed gas kinematics. We perform a global
$\chi^2$-fit of an axisymmetric rotation curve model to the 16 observed rotation curves. The
model is based on a combination of stellar and dark halo rotation curves which were 
scaled to an average rotation curve from the six slit positions closest to the
major axis (cf. Fig.~\ref{rc_combin}). Deriving \emph{i} and PA is an iterative
process. The final
values get determined only after the first full hydrodynamical simulations, when actual
non-axisymmetric model rotation curves can be fitted to the observed kinematics.
As it turns out, in the case of NGC\,4254 the difference in the resulting inclination
can be considerable. An axisymmetric model yields a disk inclination of $i = 30$\fdg8
while the simulations suggest an inclination of $i = 41$\fdg2. We use the latter
as the inclination of the galaxy. However, spatial deprojection effects are not
particularly large in the inclination range of 30\degr\ -- 40\degr. The major axis position
angle is not sensitive to the details of the rotation curve; we adopt PA = 67\fdg5.

\subsubsection{Calculation of the Gravitational Potential}
The smoothed image is then used to calculate the maximal disk stellar potential
by numerical summation over the whole mass distribution of the galaxy, which is defined
by the surface mass density $\Sigma(\mbox{\bf r})$ we get from the K-band images.
\begin{equation} \label{phidef}
\Phi_{\star}(\mbox{\bf r}_i) = - G \Upsilon_{\star} \sum_{j \neq i}
\frac{\Sigma(\mbox{\bf r}_j)}{|\mbox{\bf r}_i - \mbox{\bf r}_j|}
\end{equation}
with:
\begin{equation}
\label{deltar}
| \mbox{\bf r}_i - \mbox{\bf r}_j | = \sqrt{ \Delta {\rm x}^2 + \Delta {\rm y}^2 + \epsilon^2}.
\end{equation}
The indices \emph{i} and \emph{j} denote different radius vectors, specifying here the pixels of 
the image array. Beyond the size of the detector array, the
surface mass density is assumed to be zero. The stellar mass-to-light ratio $\Upsilon_{\star}$
is assumed to be constant, after the corrections described in \S \ref{k_light}.
The factor $\epsilon$ in Equation (\ref{deltar}) accounts for the finite thickness of galactic
disks following \citep{gne95}. This softening factor was chosen to be compatible
with a vertical exponential density scale height of $h_z = 400$ pc, independent of radius.
This assumption most probably holds for NGC 4254, which is a late type spiral and its central
spheroid, which appears barred and extends only to about a disk scale length.
Additionally, small bulges may not be spherical and thus might be well described by the constant
scale height of 400 pc assumed for most of the simulations. Moreover, our derived potential
does not depend significantly on the choice of $h_z$. Varying $h_z$ by 10\% leads to relative
changes of the resulting forces by $\sim 5.1 \times 10^{-3}$ on average. Only for the most
massive stellar disk models we performed an explicit disk-bulge decomposition and substantially
increased the softening factor for the bulge to 760 pc. This is helps to achieve a better fit
at the very inner part of the observed rotation curve.

We also calculate a potential $\Phi_{\star ,{\rm ax}}$ from an equivalent \emph{axisymmetric}
density distribution fitted to the measured K-band luminosity profile. The rotation curve is
evaluated from the stellar potential by
\begin{equation}
v^2_{\star ,{\rm ax}}(R) = R \left. \frac{d\Phi_{\star ,{\rm ax}}(r)}{dr}\right| _R.
\end{equation} 
The maximal stellar mass-to-light ratio in equation (\ref{phidef}) is determined
by fitting the rotation curve emerging from the axisymmetric potential to the inner part of the
observed rotation curve with the highest possible $\Upsilon_{\star}$. Later, when
combining stellar and dark halo potentials, we take fractions ${\rm f_d}$ of this
stellar 'maximum disk' fit, to explore various luminous-to-dark matter ratios.

\subsubsection{The Dark Halo Potential}

For the present analysis we use the radial density profile of a pseudo-isothermal sphere as the
dark matter component in our model. Its radial density profile, characterized by a
core of radius $R_c$ and a central density $\rho_c$, is given by:
\begin{equation}
\rho(R) = \rho_c \left[ 1 + \left( \frac{R}{R_c} \right)^2 \right]^{-1}.
\end{equation}
The corresponding rotation curve is
\begin{equation}
v_{\rm halo}^2(R) = v_{\infty}^2 \left[ 1 - \frac{R_c}{R} \arctan \left( \frac{R}{R_c} \right) \right]
\end{equation}
\citep{ken86} and the potential is
\begin{equation}
\Phi_{\rm halo}(R) = \int_0^R \frac{v_{\rm halo}^2(r)}{r}dr.
\end{equation}
The asymptotic circular velocity in this (infinite mass) halo $v_{\infty}$ is related to the 
two other free parameters $R_c$ and $\rho_c$ by
\begin{equation}
v_{\infty} = \sqrt{4 \pi G \rho_c R_c^2}.
\end{equation}
Thus, $v_{\infty}$ and $R_c$ uniquely specify all properties of the halo. For any fraction ${\rm f_d}$
of the maximal stellar mass we can obtain the halo parameters from the best fit of the \emph{combined}
stellar and halo rotation curve to the observed kinematics.

In the final potential the two components are combined in the following way:
\begin{equation}
\Phi_{\rm tot}(\mbox{\boldmath$R$}|{\rm f_d}) = {\rm f_d}\, \Phi_{\star}(\mbox{\boldmath$R$}) + \Phi_{\rm halo}(\mbox{\boldmath$R$}|{\rm f_d})
\end{equation}
with ${\rm f_d}$ ranging from 0 to 1, and $\Phi_{\star}$ denoting the stellar potential with maximal
$\Upsilon_{\star}$.
The contributions to the circular velocity add quadratically:
\begin{equation} \label{mldef}
v^2_{\rm tot}(\mbox{\boldmath$R$},{\rm f_d}) = {\rm f_d}\, v_{\star}^2(\mbox{\boldmath$R$}) + v_{\rm halo}^2(\mbox{\boldmath$R$},{\rm f_d}).
\end{equation}
For every ${\rm f_d}$ we are able to find a dark matter profile, which complements the luminous matter
rotation curve. $\chi^2$-fits to the observed data are similar (cf.~Table~\ref{potpar}),
necessitating this study.

\begin{deluxetable}{lccc}
\tablewidth{7.5cm}
\tablecaption{Dark halo parameters. \label{potpar}}
\tablehead{ \colhead{${\rm f_d}$} & \colhead{$R_c$} & \colhead{$v_{\infty}$} & \colhead{$\chi^2 / N$} \\
\colhead{$(M_D / M_{Dmax})$} & \colhead{(kpc)} & \colhead{$({\rm km\,s^{-1}})$} & \colhead{ }}
\startdata
0.2 & 1.08 & 155 & 2.45\\
0.4444 & 2.00 & 150 & 3.42 \\
0.6 & 3.03 & 150 & 3.30\\
0.85 & 5.68 & 155 & 2.00\\
1.0 & 7.30 & 155 & 1.84\\
\enddata
\tablecomments{Dark halo parameters used to generate the potentials used for simulations.
The $\chi^2 / N$-values refer to an axisymmetric model.}
\end{deluxetable}

\subsubsection{The Pattern Speed $\Omega_{\rm p}$ and Corotation}
The spiral structure in galaxies is tightly related to the density wave theory that is used to 
describe the formation of these structures in spiral disks
\citep{ath84}. If the spiral pattern evolves more slowly
than the orbital time scale, resonances occur at certain radii in the galactic disk. 
We assume that the spiral pattern remains constant in
a particular corotating inertial frame, specified by
the pattern speed $\Omega_{\rm p}$. At the corotation radius ${\rm R}_{\rm CR}$
($\Omega_{\rm p} = {\rm V} / {\rm R}_{\rm CR}$ where V is the circular rotation speed) everything
corotates with the spiral pattern. In addition to this corotation resonance there are more
radii, at which the radial oscillation frequency $\kappa$ and the circular frequency $\Omega$
are commensurate. In linear density wave and modal theories, these so called inner
and outer Lindblad resonances should inclose the radial range for which a $m = 2$ spiral pattern
may develop and persist over a longer period of time \citep{lin64}. On the other hand, non-linear
orbital models for open galaxies indicate that the symmetric, strong part of the stellar
spiral might end at the inner 4:1 resonance \citep{pat91}.
Finding the radial locations of these resonances in real galaxies is a delicate endeavor and
there are no widely agreed upon ways to determine them.

For our analysis the pattern speed $\Omega_{\rm p}$ or its corresponding corotation radius, is
crucial since it modifies the forces acting in a non-rotating inertial frame
\begin{equation}
\mbox{\bf F}(\mbox{\boldmath$R$}) = \mbox{\boldmath$\nabla$}\, \Phi_{\rm tot}(\mbox{\boldmath$R$}) + \Omega_{\rm p}\, R^2.
\end{equation}
Therefore we need to perform a set of hydrodynamical simulations at different $\Omega_{\rm p}$, to
find the best matching value and thereby to also determine the corotation radius. As it turns out,
the simulations are well suited to find these values reliably. However, to reproduce all 
the gravitationally induced features found in the observed kinematics, we suspect that the assumption
of a constant pattern speed for the whole disk is only a crude approximation of the
dynamical processes occurring in the disk of NGC 4254. The pattern rotation speed most
probably depends slightly on the radius. A radius dependent pattern speed would eventually drive
evolution of the spiral pattern. Although spiral structure is most likely a transient feature
in the galactic disk, it persists for several dynamical time scales \citep{don94,pat99}
\footnote{In light of our project it is interesting to note that in both N-body simulations
referenced here, sub-maximal disks were adopted in the initial conditions}.
But, since for now we have no means to reliably determine the
radial dependence of the pattern rotation speed, we choose to ignore the time evolution of the
spiral pattern and assume a constant pattern speed for the simulations.

\subsection{Hydrodynamical Simulations} \label{hydrosim}
The next step is to calculate the expected kinematics for a cold gas orbiting in $\Phi_{\rm tot}$
with a certain $\Omega_{\rm p}$. We model the gas flow with the BGK hydro-code
\citep{pre93,sly99}. This is an Eulerian,
grid-based hydrodynamics code which is derived from gas kinetic theory. The simulations
are challenging because cold gas in galactic disks rotates with such high Mach numbers
that if the flow is diverted from a circular orbit by non-axisymmetric forces, the
consequence is etched in the gas in patterns of shocks and rarefied regions. BGK is well
suited for these simulations because as verified by extensive
tests on standard 1D and 2D test cases of discontinuous non-equilibrium flow
\citep{xu98}, at shocks and contact discontinuities BGK behaves
as well as the best high resolution codes and it gives better results at rarefaction
waves because it naturally satisfies the entropy condition.

To model the two-dimensional gas surface densities and velocity fields for NGC 4254 we
carried out a set of simulations on a 201 by 201 evenly spaced Cartesian grid.
Our data for NGC 4254 extend out to a radius of about 11.6\,kpc,
hence on a 201 by 201 grid this gives a resolution of about 116\,pc per side of
a grid cell, which is considerably higher than the effective force smoothing of 400\,pc.

Given that the gas surface mass density of the modelled galaxies is much lower
than the density of the stellar disk and halo, we neglect the
self-gravity of the gas, and compute its response to a fixed
non-axisymmetric gravitational potential derived from the corrected NIR
image of the galaxy with a 3-fold reduced resolution (cf. Sect.~\ref{modstarpot}).
We approximate the ISM as a monatomic gas.

The gas temperature profile is taken to be uniform and by imposing an isothermal equation of
state throughout the simulation we assume that the gas instantaneously cools
to its initial temperature during each updating timestep.
The initial density profile of the gas is exponential with a scale length equal
to a third of the disk radius, namely 3.86 kpc. We begin each simulation with the
gas initially in inviscid centrifugal equilibrium in the axisymmetric potential
given by $\Phi_{\star ,{\rm ax}}$ and $\Phi_{\rm halo}$ (see Sect.~\ref{modstarpot}).
Following the initialization of the gas in centrifugal
balance in an axisymmetric potential, we slowly turn on the non-axisymmetric
potential at a linear rate computed by
interpolating between the final non-axisymmetric potential and
the initial axisymmetric potential so that the potential is fully
turned on by the time 40 sound crossing times of the code have passed.
Here the sound crossing time is defined as the time it takes to traverse the
length of the diagonal of a grid cell at sound speed. For an isothermal
simulation with a sound speed of $10\, {\rm km\,s}^{-1}$ the sound crossing time
of a cell in our simulation is about 16 Myr, so that by 40 sound crossing
times, the gas has evolved for 640 Myr which is about 1.4 times the dynamical
time of the galaxy measured at a radius of 11.6 kpc. After the non-axisymmetric
part of the gravitational potential has been fully turned on, we continue to
run the simulation for about another dynamical time. A thorough description of
the technical details of the simulations is given in a companion paper (Slyz et al. (2001), in prep.).

We ran a large set of simulations both to understand the power and limitation of our modelling
in general and to match the observations.
Simulations were performed for a total of five different fractions ${\rm f_d}$
of the stellar disk: disk only, i.e.~${\rm f_d} = 1$, and ${\rm f_d} = $ 0.85, 0.6, 0.4444 and 0.2,
or accordingly ${\rm f_d}$ is given in \% from 20\% to 100\%. In all the cases the core radius and the
asymptotic velocity of the pseudo-isothermal halo were adjusted to best match the averaged
rotation curve, as summarized in Table \ref{potpar}.
The variations in $\chi^2 / N$ between the low mass disks and the massive
ones are mainly caused by the attempt to keep $R_c$ and $v_{\infty}$ at physically
reasonable values. The bump at 20\arcsec\ to 40\arcsec\ (Fig.~\ref{rc_combin})
is fitted better for the high mass disks, which reduces the overall $\chi^2$
compared to the low mass disks.

We have no secure prior knowledge of the spiral pattern speed
$\Omega_{\rm p}$. We determine it by assuming different values for $\Omega_{\rm p}$
and then comparing a simulation to the data. For every simulation with a different
stellar/dark halo combination, we get slightly different values for
the best matching $\Omega_{\rm p}$ or equivalently for the corotation radius
$R_{\rm CR}$.
We covered the complete range of reasonable $R_{\rm CR}$, i.e.~from about a disk
scale length to well outside the disk. We even made 
simulations for the case of no spiral pattern rotation, $R_{\rm CR} \rightarrow
\infty$.

To test how the amplitude of the velocity perturbations depends on the responsiveness
of the gas, we ran simulations at a variety of temperatures corresponding to sound
speeds $c_{s}$, of $10,\,20,\,30\, {\rm and}\, 40\, {\rm km\,s}^{-1}$.
In the following Sections we discuss some of the results from the simulations.

\section{RESULTS FOR NGC 4254}
\subsection{Simulated Gas Density} \label{gasdensity}

\begin{figure*}[t]
 \resizebox{\hsize}{!}{\includegraphics{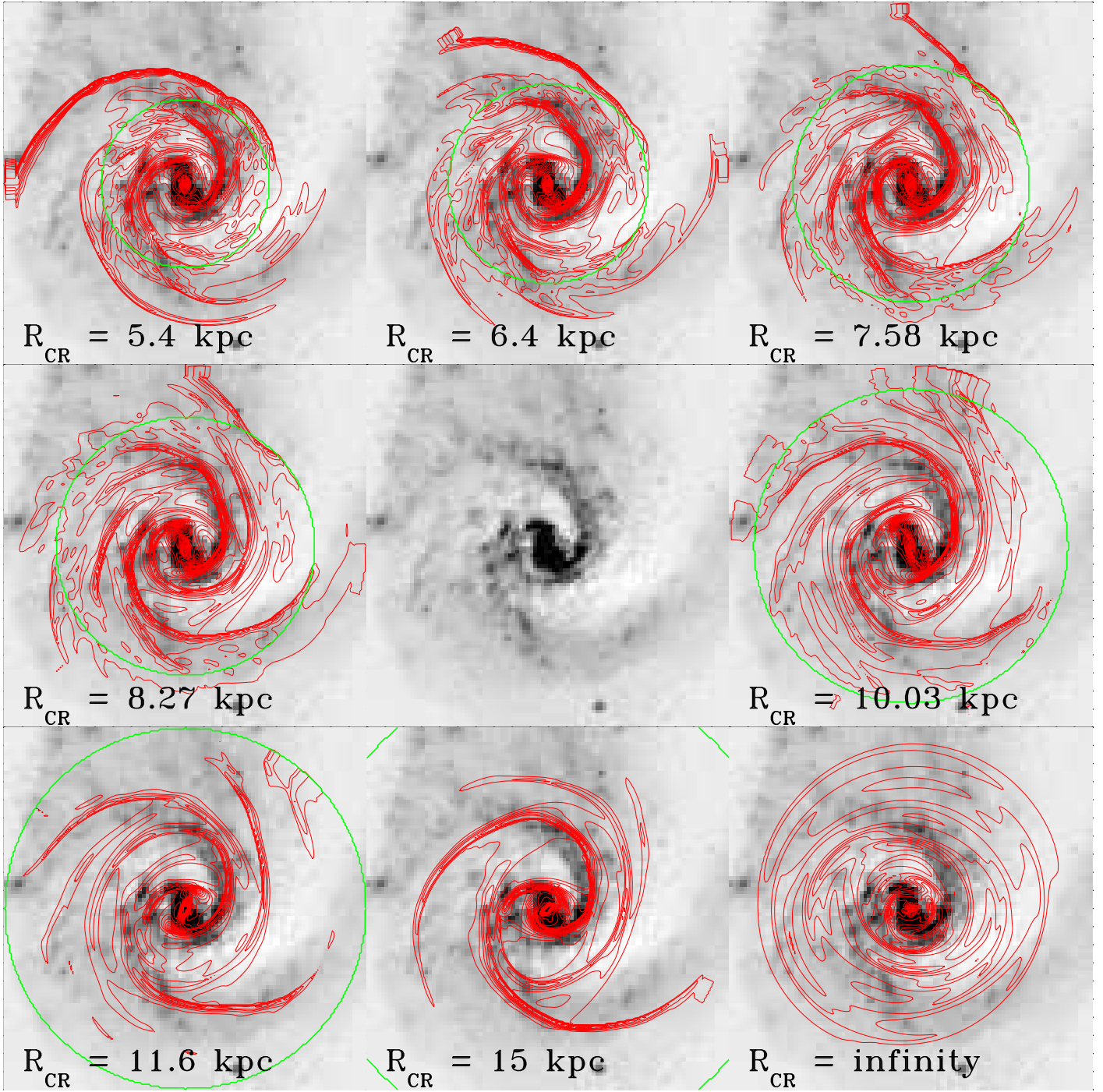}}
 \caption{Simulation results of the gas density distribution overplotted in red contours on
 the deprojected K'-band image of NGC 4254 (center). From the image an axisymmetric radial
 brightness profile has been subtracted to enhance the contrast of the spiral arms. The results
 are displayed for eight different assumptions for the pattern speed, respectively the corotation
 radius $R_{\rm CR}$. The green circle marks corotation. The range goes from fast rotation
 $R_{\rm CR}$ = 5.4 kpc to slow pattern rotation $R_{\rm CR}$ = 15 kpc, lying mostly outside the
 image frame and finally for no pattern rotation.
 \label{simgasdens}}
\end{figure*}

Figure \ref{simgasdens} shows eight views of the simulated gas density for different pattern speeds
over-plotted as contours over the unsmoothed, deprojected, color corrected K-band image of NGC 4254.
The simulated gas density follows an overall exponential profile with a scale length of
$\approx 4.2$ kpc, comparable to the one of the disk itself. The contours in the Figure are chosen to
highlight the density enhancements and locations of the gas shocks caused by the spiral
arms. For almost all simulated cases the strong part of the galaxy's spiral structure
lies well inside corotation, where the circular velocity is larger than the
spiral pattern speed. The gas will thus enter the spiral
arms from their inward facing side, producing the strongest shocks there. For a well matching 
simulation, we expect the shocks to be near the OB associations that trace the spiral
arms.

It is remarkable how well the overall morphology of NGC 4254 can be
matched by the gas density simulations. Not only are the two major spiral arms clearly identifiable 
in most simulations but the less prominent northern arm and the locations where the arms bifurcate
are reproduced in some cases well. For fast pattern speeds ($R_{\rm CR} = 5.4\, -\, 7.58\, {\rm kpc}$ in
Fig.~\ref{simgasdens}) we find a strong shock in the northern part of the galaxy that cannot be
correlated with any mass feature. We believe it develops because the potential close to the upper
boundaries of the computational grid is quite non-axisymmetric,
and this leads to a spurious enhancement of a shock. The shock does not propagate into regions
inside the corotation radius, and therefore we refrain from smoothing the potential.

It is important to note that the simulations lead to a very stable
gas density distribution that does not change much after the non-axisymmetric
potential is fully turned on.
When the contribution of the disk is increased in the combined potential, all spiral
features get enhanced in the gas density but the galaxy morphology is essentially unchanged.
For a more detailed discussion of these issues, please refer to the
accompanying paper by Slyz et al., which is in preparation. With increasing pattern rotation
(smaller corotation radii in Fig.~\ref{simgasdens}) we find
that the predicted spiral arms become more and more tightly wound.
 For a comparison to the
stellar spiral morphology we need to define some criteria to pick the right model.
If the situation in NGC 4254 is similar to NGC 4321, whose gas and dust
distributions and their connection to star forming regions have been
discussed in detail by \citet{kna96}, then a good matching
gas morphology is one where for radii smaller than the corotation
radius, the shocks in the gas density lie on the inside of the stellar
spirals. Shortly downstream from there, many star forming \ion{H}{2} regions,
triggered by gas compressions, should show up as patches in the arms.
According to these criteria, the best matching morphology
 
can be unambiguously identified to be produced by a simulation with
$R_{\rm CR} \approx 7.6\, {\rm kpc}$ (Fig.~\ref{simgasdens} upper
right panel), corresponding to a pattern speed of
$\Omega_{\rm p} \approx 20\, {\rm km\,s}^{-1}{\rm kpc}^{-1}$. $R_{\rm CR} = 6.4\, {\rm kpc}$ and
$R_{\rm CR} = 8.3\, {\rm kpc}$ enclose the range of possible values. This corresponds to
an uncertainty of $\sim \, 15$\% in the value of $R_{\rm CR}$.

 Our results were
compared to values of $R_{\rm CR}$ for NGC~4254 from the literature, which were determined
by different means. The results $R_{\rm CR} \sim 8.45\,$ kpc \citep{elm92} and
$R_{\rm CR} = 10.2 \pm 0.8$ kpc \citep{gon96}, scaled to our distance assumptions,
provide larger estimates than our findings. However, given the picture of non-linear orbital models,
where one expects the strong part of the stellar spiral to end inside the inner 4:1 resonance
\citep{pat91} we find that our result of $R_{\rm CR} \approx 7.6\, {\rm kpc}$ is consistent with
the galaxy's morphology. 
In short, these simulated gas densities provide us with an excellent tool to determine
the pattern rotation speed of the galaxy. Apart from requiring a constant global pattern rotation
our approach is independent of an underlying spiral density wave model.
The overall very good representation of the whole
spiral structure by the simulated gas density makes us rather confident that the simulations
render realistic processes affecting the gas.

\subsection{Simulated Gas Velocity Fields}

As another output of our simulations we get the two-dimensional velocity field of the gas.
As is evident from a comparison
to the gas density distribution, the velocity jumps are -- as expected -- at the locations 
where the density map shows the
shocks. They show up as areas of lower local circular velocity compared to the elsewhere rather
smoothly varying gas velocity field. The velocity wiggles, as well as the density shocks
themselves, have very tight profiles and thus are even more narrow than the physical extent
of the stellar arms. They have to be compared to the observed kinematics.

\subsubsection{The observed kinematics}
The rotation curves at the 16 slit positions from the observations are shown in
Fig.~\ref{allslits} as data points. It is apparent that the longslit spectra allow a good
velocity coverage along the slits. Almost all rotation curves show contiguous data points out
to a radius of $\ga 1$\farcm5. The spectra show a lot of wiggles on a small spatial scale.
Jumps of $\la 30\, {\rm km\,s}^{-1}$ on a scale of $\sim 5$\arcsec\ are common. The very
prominent jumps that we observed in slit positions 22\fdg5 and 225\degr\ clearly exceed
the average wiggle sizes. Inside of about 0\farcm3 the small bar influences the velocity field.
The most prominent trace of the bar occurs at its minor axis at the slit positions of
135\degr/157\fdg5 and 315\degr/337\fdg5.

The trace of kinematic features in the outer disk is not conspicuous in subsequent slits.
The eastern part of the disk (slit positions 45\degr\ -- 135\degr) displays a quite smooth
velocity field, while the western part shows some large scale variations. Aside from the
inter-arm region between the inner disk and the southern arm where a $\approx 0$\farcm5
wide depression is moving outward in subsequent slits (positions $\ge 247$\fdg5) no significant
features are apparent in the outer disk. Unfortunately in this inter-arm region the $S/N$
is not so good. Does that mean we do not see the trace of the arms in the velocity field, or
is the single slit representation of the 2D velocity field misleading and does not allow us to
identify coherent features in adjacent slits? Clearly, the wiggles associated with spiral arms
in NGC 4254 are not nearly as strong as in M81 \citep{vis80}, thus their identification is harder.
A CO map of NGC 4254's center \citep{sak99} shows also no coherent wiggles across spiral arms
and we doubt that it would be much different on a Fabry-Perot image. Rather than being confused
by the one dimensional nature of our rotation curve slices we believe that the spiral features in
the velocity field are intrinsically weak.

\subsubsection{Projection and Alignment}
To compare the simulated gas velocities to the observed data, we need to project the modelled
velocity field according to the real orientation of the galaxy
so that we can obtain the line-of-sight velocity along locations corresponding
to the slit positions taken with the spectrograph. For this procedure we take the velocity
components and transform them into velocity components parallel
and orthogonal to NGC 4254's major axis. Since the simulations yield truly 2D
velocity fields, the component parallel to the major axis does not contribute to the
line-of-sight velocity as it reflects only tangential motion. The orthogonal component is
multiplied by the sine of the inclination of the disk, to account for the line-of-sight
fraction of the velocities.

We read the velocities along slits which correspond to the slit positions along which we
took measurements with the spectrograph. The angular width of the grid cells of the simulation
(1\farcs 19) is comparable to the slit width of the spectrograph we used (1\farcs 5), which
was also the average seeing conditions. 

The angles between the spectrograph slit orientations must be
translated to angles in the plane of the galaxy to actually compare the same parts of the 
velocity fields. This translation makes use of the following relationship:
\begin{equation}
\tan(\varphi_{\rm int} - {\rm PA}) \cdot \cos(i) = \tan(\varphi_{\rm app} - {\rm PA})
\end{equation}
where $\varphi_{\rm app}$ is the apparent angle of the spectrograph slit across the galaxy 
on the sky and $\varphi_{\rm int}$ is the corresponding intrinsic angle within the plane of
the galaxy. PA is the position angle of the galaxy's major axis on the sky. All angles
are in degrees measured eastward from north. Solving for $\varphi_{\rm int}$ gives:
\begin{equation}
\varphi_{\rm int} = \arctan \left( \frac{\tan(\varphi_{\rm app} - {\rm PA})}{\cos(i)} \right) + {\rm PA}.
\end{equation}
The foreshortening of the radial proportions due to projection in the direction of the
minor axis is determined by
\begin{equation}
R_{\rm proj} = R \cdot \sqrt{\cos^2(\varphi_{\rm int}) + (\sin(\varphi_{\rm int}) \cdot \cos(i))^2}.
\end{equation}
Finally the detector pixel sizes of the TWIN, where the observed velocities come from, and Omega Prime
camera, whose pixel scale is the reference grid for the simulations, must be adjusted to perfectly align with
each other. 

\subsubsection{Overall Fit Quality} \label{fitquality}

\begin{figure*}[t]
 \resizebox{\hsize}{!}{\includegraphics{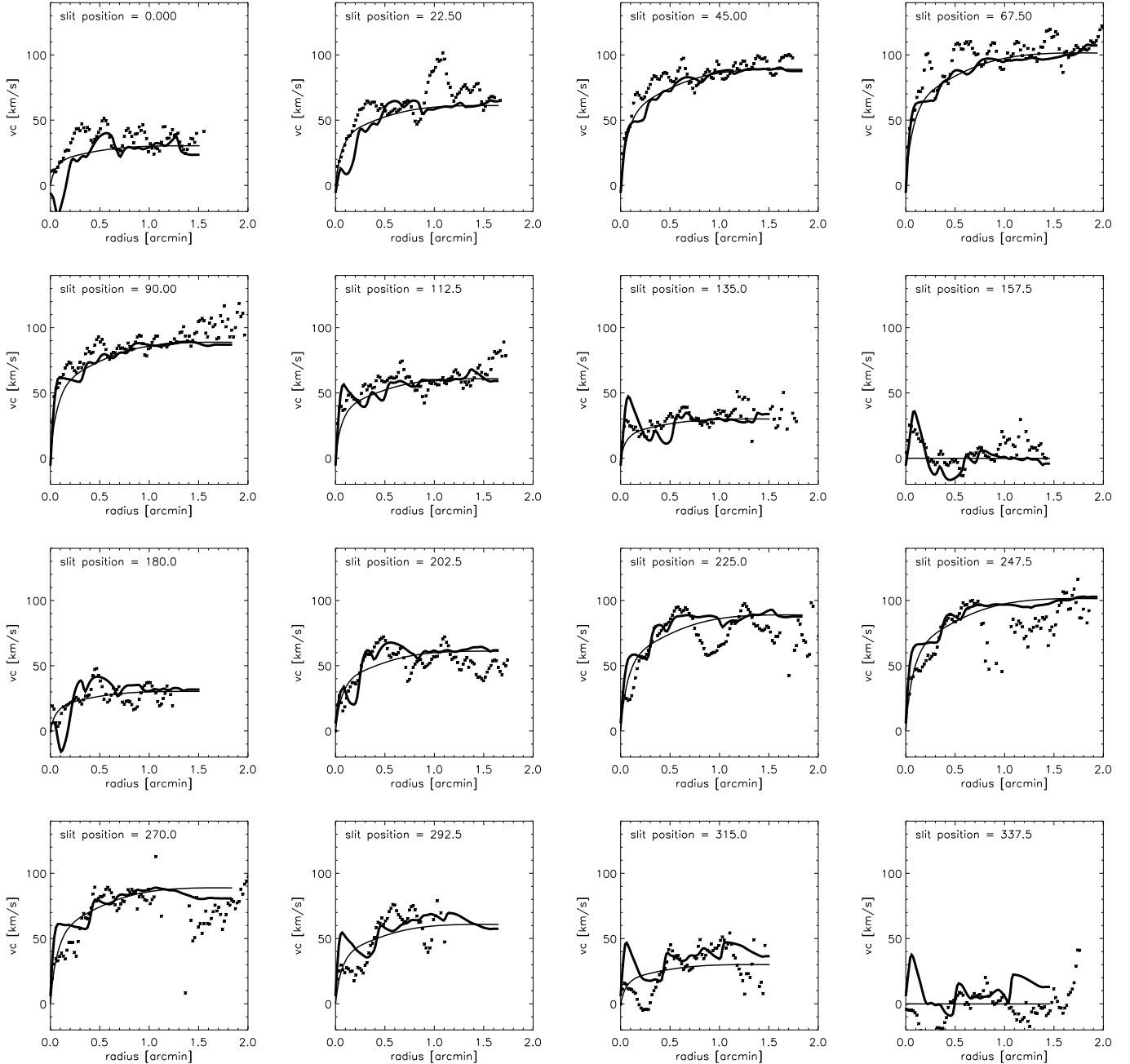}}
 \caption{Simulation results of the gas velocity field in comparison to the observed rotation
curves. Displayed are the measured rotation curves (data points), the axisymmetric model
(thin line) and the rotation curves from the hydrodynamical simulations (thick line)
for all 16 slit position angles. The parameters for the simulation were:
${\rm f_d} $ of 60\% and $R_{\rm CR} = 7.58$ kpc. There are no error bars plotted for the data,
but the errors can be estimated from the point-to-point scatter of the data.
 \label{allslits}}
\end{figure*}

Figure \ref{allslits} shows the 16 separate rotation curves with a corresponding 
simulated velocity field over-plotted. The simulation used here for comparison
is the one for which the gas density distribution best-fit the K-band image (displayed in
Fig.~\ref{simgasdens}). It has a corotation radius of $R_{\rm CR} = 7.58$ kpc --
corresponding to a pattern speed of $\Omega_{\rm p} = 20\, {\rm km\,s}^{-1}{\rm kpc}^{-1}$
-- and a stellar disk mass fraction ${\rm f_d} $ of 60\%. A gas sound speed of
$c_s = 10\, {\rm km\,s}^{-1}$ was assumed here.

An overall fit quality gets determined by a global $\chi^2$-comparison of the simulated
velocity field to the actual observed velocity field along the 16 measured slit positions. Since
with the errors from the observations we find a $\chi^2/N > 1.0$ we added an overall
$9.5\, {\rm km\,s}^{-1}$ to provide an average fit with $\chi^2/N \approx 1.0$. The $\chi^2$-fitting
excludes the very central region hosting the small bar because the modelling is not
intended to fit the central bar, which might have a different pattern speed. The total number of
data points included in the $\chi^2$-fitting is 1077.

The general fit quality is governed by the effect 
that the projection of the simulated velocity field introduces. The good overall match indicates
that we quite reliably found the right position angle and inclination for the galaxy.
The simulated velocities align very well with the measured data
points. In addition to the good overall match, the general rising or falling shape of the separate
curves is also excellently reproduced by the simulations. The lopsidedness of the galaxy is reflected
in the shape of the rotation curves on the receding and approaching side of the disk.
At the receding side (67\fdg5) the rotation curve rises steeply and continues to rise 
out to 2\arcmin, while the approaching side (247\fdg5) rises less steeply but flattens out or even
drops beyond 0\farcm 7. These characteristics are closely reproduced by the models.

A close inspection of the two profiles shows however that the overlap in the match
of the simulated velocities with the measurements is not always satisfactory.
The agreement of local features in the simulations and the measured data is sometimes 
very good and even occurs in subsequent slit positions. However, there are also many
locations, where the match is poor. This is particularly the case in the inner region of the
galaxy, where the small bar dominates the kinematics. Both profiles show strong wiggles
where the slit crosses the bar, especially close to the minor axis of NGC 4254's
velocity field, which is also close to the minor axis of the bar itself. While the 
simulations show a rather symmetric imprint, the measurements exhibit a signature different
from that, leading to a significant mismatch at several slit positions, e.g. 292\fdg5 and
337\fdg5. This might be caused by the bar, having a slightly different pattern speed.
In the outer parts of the rotation curve we also find several wiggles in the observed
data that have no correspondence to the wiggles in the simulations and vice versa.

\begin{figure*}[t]
 \resizebox{\hsize}{!}{\includegraphics{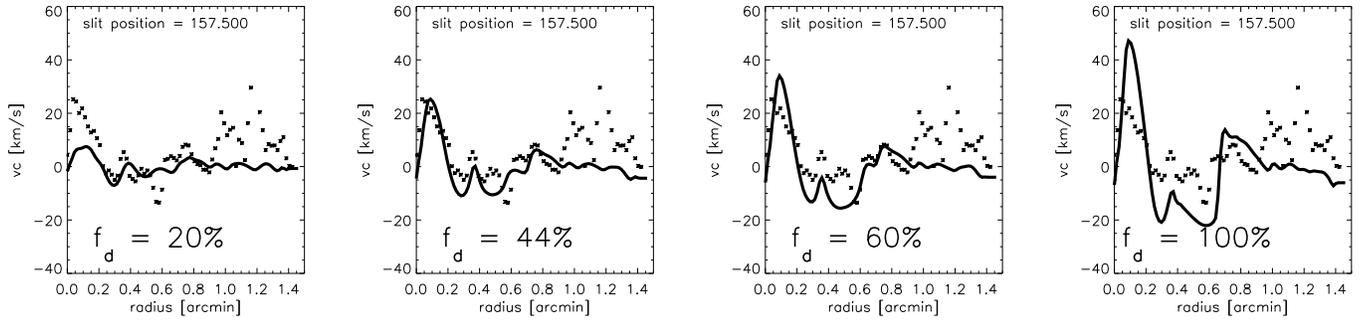}}
 \caption{Comparison of four simulations for different fractions ${\rm f_d}$ of the stellar
mass component. The simulation was done with a corotation radius of 7.58 kpc and a
gas sound speed of $10\,{\rm km\, s}^{-1}$. Displayed is one of the 16 slit positions.
Clearly the predicted ''wiggles'' in the rotation curve grow much stronger for higher disk
mass fractions.
 \label{mlcompare}}
\end{figure*}

It is important to note that we do not expect to reproduce all the wiggles
in the galaxy's rotation curves, since we are only modelling those created by the non-axisymmetric
gravitational potential. The wiggles originating for another reason -- like expanding SN gas shells --
are not considered by the simulations and thus do not show up in the resulting velocity field.

One very interesting thing to mention is the fact that even for moderate differences in the fit
quality of the axisymmetric disk model (cf.~Table \ref{potpar}), the fit quality of the non-axisymmetric
simulated velocity field is rather independent of the initial axisymmetric model.
And, moreover, the formally preferred axisymmetric maximum-disk decomposition (cf.~Table \ref{potpar})
turns out to be the most unfavored model, once the simulations were performed.
This implies that even if an axisymmetric model profile provides a better fit to a measured rotation
curve, it does not necessarily mean, that this combination provides the best fit when one considers
the 2D non-axisymmetric gas evolution.

\subsubsection{Varying the Stellar to Dark Matter Ratio}
\begin{figure}[h]
 \resizebox{\hsize}{!}{\includegraphics{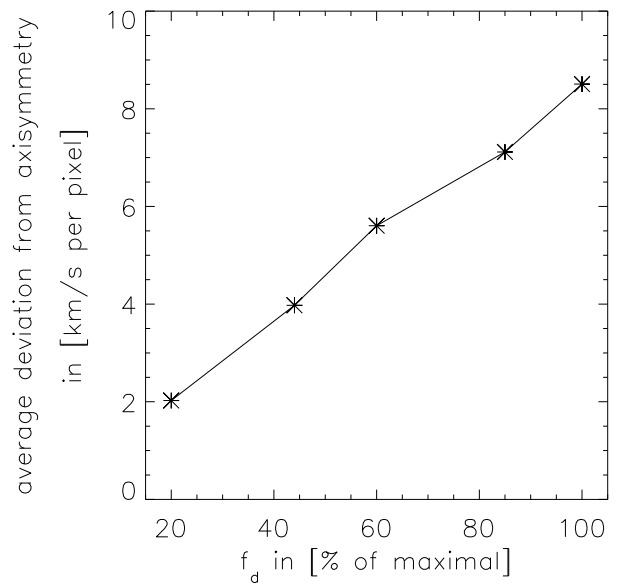}}
 \caption{Deviation of the simulation from axisymmetry. Displayed is the average deviation
of each radial simulation bin from the axisymmetric rotation curve. It rises linearly
with the stellar disk mass contribution ${\rm f_d}$.
 \label{linwigg}}
\end{figure}

As already mentioned in Section \ref{hydrosim}, we performed simulations for five stellar disk
and dark halo combinations, as listed in Table \ref{potpar}. Since the non-axisymmetric
perturbations are induced in the potential by the stellar contribution, we expect the
amplitude of the wiggles in the modelled rotation curves to depend significantly on the
non-axisymmetric contribution of the stellar potential whereas we expect the radial
distribution of the wiggles to be rather independent of the stellar mass fraction.
As expected, in the simulations with the lightest disk, the wiggles look like modulations
on the axisymmetric rotation curve. In the case of the maximum disk, the rotation curves
are strongly non-axisymmetric (cf. Fig.~\ref{mlcompare}). To describe this 
characteristic more quantitatively, we learn from Fig.~\ref{linwigg} that
the amplitude of the deviations from axisymmetry increases linearly with the mass
fraction of the stellar disk, which proves the general validity of the concept with which
we try to approach this problem.

The strongest velocity wiggles arising in the modelled velocity fields are the ones
connected to the central bar-like feature. However, since here we are not interested in
modelling the dynamics of the bar, we exclude this inner part from the analysis.

\begin{figure}[h]
 \resizebox{\hsize}{!}{\includegraphics{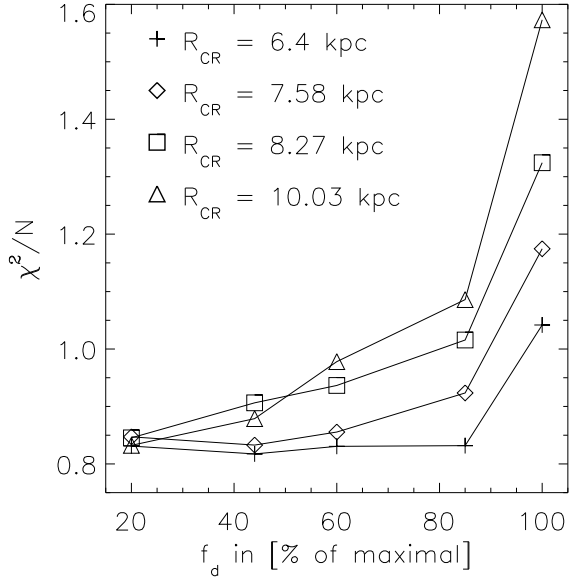}}
 \caption{Formal $\chi^2$-fit of the gas velocity simulations for different stellar mass
 contributions ${\rm f_d}$, normalized by the $\chi^2$-fit for the axisymmetric model
 rotation curve. Displayed are cases for four different values of $R_{\rm CR}$.
 From the simulations we can rule out the cases with the highest disk mass.
 \label{chi24fds}}
\end{figure}

If we do a formal $\chi^2$ comparison of the models with the observed data we find that for
most of the sub-maximal disks the fit is considerably better than for heavy stellar
disks. By formal, we mean that we use all data points for the $\chi^2$-fit, regardless
of whether a certain part matches well or not with two exceptions: we exclude the very
central part with the bar and we correct for the outer strong shock appearing in the fast
rotating models (cf.~Fig.~\ref{simgasdens}). 

The result 
from this $\chi^2$-fit is presented in Fig.~\ref{chi24fds}. In all cases, ${\rm f_d} = 100$\%
gives the worst fit to the observed rotation curves. For the lower mass disks it is 
very hard to decide, whether a particular disk mass is preferred. For $R_{\rm CR} = 6.4$
kpc we find about the same $\chi^2$ for all non-maximal disk models. Since we can not reject
data on a physical basis, we can state only a trend at this point. Thus our conclusion
from this part of the analysis is that the disk is most likely less than 85\% maximal.

\subsubsection{Varying the Gas Temperature} 
This kind of analysis may be very sensitive to the temperature of the gas which is assumed for 
the modelling. Higher gas temperature corresponds to a higher cloud velocity dispersion $c_s$ and 
thus to a reduction in the response of the gas to any feature in the gravitational potential. 
Within the Milky Way Galaxy $c_s$ varies from $\approx 6\, {\rm km\,s}^{-1}$ in the solar
neighborhood to $\approx 25\, {\rm km\,s}^{-1}$ in the Galactic center \citep{eng99}.
For simulations of a galactic disk with an isothermal equation of state,
the most commonly used values for $c_s$ are 8 -- $10\, {\rm km\,s}^{-1}$,
corresponding to $< 10^4$ K in gas temperature \citep{eng99,wei01b}.
In these simulations the authors make the statement that within the reasonable
limits of $c_s = 5 - 30\, {\rm km\,s}^{-1}$ the modelled gas flows across the primary shocks 
is not considerably affected.
 Only when modelling strong bars in galaxies the simulations might be
dependent on the choice of the gas sound speed \citep{eng97}.  
For the main set of simulations presented in this paper, we chose $c_s = 10\, {\rm km\,s}^{-1}$,
corresponding to a gas temperature of 7250 K. To probe the effect of the gas temperature, we performed
simulations for four different gas sound speeds, $c_s$, of 15, 20, 30 and 40 ${\rm km\,s}^{-1}$.
With a simulation performed at higher sound speed, we tested if our results at lower sound speeds
were caused by too cold gas being too strongly affected by the non-axisymmetric part of the potential.
Increasing the sound speed slightly broadened and shifted the shock fronts in our simulations,
but not enough to change the conclusions of this paper. A more detailed
technical discussion on the simulated gas dynamics in NGC 4254 will be presented in a forthcoming paper
by Slyz et al. 

\section{DISCUSSION}
\label{discussion}
To learn about the relative role of stellar and dark matter in the inner parts of
spiral galaxies, we have extended the hydrodynamical simulations of galactic features
from bars \citep{wei01b} to the regime of spiral arms. In this paper
we have used color corrected K-band images to derive the stellar portion of the
potential of NGC 4254 for the analysis.

From the simulation of the gas density we find a very good morphological match of the
gas shocks to the spiral arms in the galaxy; this aspect of the simulations
determines the pattern speed $\Omega_{\rm p}$ of the galaxy to about 15\% precision.

 A comparison of the observed and simulated kinematics has
turned out to be challenging. Although the overall shapes of the different rotation curves
were very well reproduced by the simulations, some small scale structure remains unmatched. 
The formal comparison of the gas velocity field to the observed H$\alpha$ kinematics favors
simulations with small disk mass fractions ${\rm f_d}$ (cf. Fig.~\ref{chi24fds}) and
correspondingly small values for the stellar mass-to-light ratio. 
With the K-band mass-to-light ratio discussed in Section \ref{k_light} our results yield an
overall stellar M/L of $\Upsilon_{\star} \la 0.5$. We can estimate the relative mass fractions
from their contributions
to the total rotation velocity. At a radius of 2.2 uncorrected K'-band exponential disk scale lengths
$(2.2\,R_{\rm exp} \approx 79$\arcsec\ or 7.7 kpc), the individual rotational support of the stellar
and dark halo components for ${\rm f_d} = 85$\% are $v_{\star} = 125\, {\rm km\,s}^{-1}$ and
$v_{\rm halo} = 86\, {\rm km\,s}^{-1}$. If a total mass is estimated via
\begin{equation}
{\rm M}(2.2R_{\rm exp}) = \frac{v^2(2.2R_{\rm exp})}{2.2R_{\rm exp}G}
\end{equation}
we find that $M_{\rm halo} \approx 0.47\, M_{\star}$ at $R = 2.2\, R_{\rm exp}$, or accordingly
$\ga 1/3$ of the total mass inside $R_{\rm exp}$ is dark. Since our confidence limits are not
very tight, we cannot use them to test other authors' findings in detail. Projects yielding results
in favor of a sub-maximal stellar disk usually find a disk mass fraction less than our upper limit
estimate. \citet{bot97} as well as \citet{CR99} conclude that the contribution of the stellar
disk to the total rotation is $v_{\star} \sim 0.6\, v_{\rm tot}$ which translates to
$M_{\rm halo} \sim 0.6\, M_{\rm tot}$. At the current state of the project we cannot exclude or confirm
these findings. This issue is going to be discussed more thoroughly as soon
as we have a few more examples analyzed. 

In the following Sections we will discuss the details that could cause deviations from a perfect match
between the simulations and the measurements.

\subsection{Is the Concept Reasonable?}
If the self-gravity of the stellar mass in the disks of spiral galaxies plays an important
role then undoubtedly the potential becomes non-axisymmetric. The trajectory of any kinematic
tracer in the galaxy, such as the gas, should be affected by these potential modulations.
But is the \ion{H}{2} component of the gas the best choice for tracing the galaxy's potential?
Analytic calculations of gas shocks in the gravitational potential of a spiral galaxy
\citep{rob69} tell us that we should expect velocity wiggles with
an amplitude of 10 to 30 ${\rm km\,s}^{-1}$ while crossing massive spiral arms. However, kinematic
feedback to the gas from regions of massive star formation, from expanding gas shells produced by
supernova explosions and from other sources of turbulence, introduces small-scale random noise
in the velocity fields. These fluctuations typically lie in the range of 10 to 15 ${\rm km\,s}^{-1}$
\citep{bea99} and seem even higher in the case of NGC 4254.
The kinematic small-scale noise could be increased if the dynamics of the brightest
\ion{H}{2} regions is kinematically decoupled from the global ionized gas distribution. To check
this we over-plotted the H$\alpha$-intensity on the rotation curves to see if the star formation
regions coincide with the strongest wiggles in the rotation curves. There is, however, no discernible
relation between the amplitude of a wiggle and the intensity of the \ion{H}{2} region,
indicating that the deviations are not confined to compact \ion{H}{2} regions.

As an alternative to observing the ionized phase of the hydrogen gas, one could consider using \ion{H}{1} radio
observations. However, available \ion{H}{1} data are limited by the larger size of the radio beam
that smears out kinematic small-scale structures in the gas. In principle, stellar
absorption spectra could also provide relevant kinematic information, but this approach has
two disadvantages compared to an approach using gas
kinematics. First, the acquisition of stellar absorption spectra with sufficient S/N would take
a prohibitively large amount of telescope time. Second, stellar kinematics cannot be uniquely
mapped to a given potential; there are different sets of orbits resulting in the same observed
surface mass distribution and kinematics. Thus, despite its apparent shortcomings \ion{H}{2} measurements
seem to be the most promising method with which to approach the problem.

Could the discrepancies between modelled and observed kinematics be related to taking
the NIR K-band image of a galaxy as the basis to calculate its stellar potential? As already
discussed in Section \ref{modstarpot} there are several factors which throw into question whether
the K-band image is a good constant mass-to-light ratio map of the stellar mass distribution despite
the color correction we applied. However, the two major factors -- dust extinction and the
population of red super giant stars -- tend to affect the arm-interarm contrast, rather than the 
location of the K-light spiral arms. Dust lanes lie preferentially inside the $m = 2$ component of a
galaxy's spiral \citep{gro00} absorbing interarm light, while the red super
giants may actually have their highest density in the spiral arms directly, where they emerged
from the fastest evolving OB stars. This effect should become apparent in the simulations 
as slightly wrong amplitudes of the gas wiggles, leaving their radial position mostly unchanged.
So even if the K-band images might actually include unaccounted mass-to-light ratio variations,
they most probably introduce only small errors, which should not result in an overall
mismatch of the models with the data. The color-corrected K-band images should therefore reflect
the stellar mass accurately enough for our analysis. Certainly, there appears to be no better
practical mass estimate that we could use for our analysis.

\subsection{Are there Systematic Errors in the Modelling?}
The most critical part of this study are the several modelling steps required
to predict the gas velocity field for the comparison with the data. We apply a spatial
filter to the K-band image before calculating the disk potential to reduce the 
significance of the clumpy \ion{H}{2} regions (as described in Sect.~\ref{modstarpot}).
The residual map of the discarded 
component shows all the bright \ion{H}{2} regions in the disk, giving us confidence that we have 
excluded much of the structures that reflect small-scale M/L variations. 
This correction removes roughly 3.5\,\% of the total K-band light and does not depend strongly
on the number of Fourier components used for the fit. For two extreme decompositions employing
6 and 16 Fourier components, the mean resulting relative discrepancy in the derived potentials is
only $\approx 10^{-6}$.  
One remaining concern in calculating the potential might be that the mass density map cuts at
the border of the image. However, the galaxy completely fits in the frame, fading
into noise before the image border. Moreover, we do not perform the simulations for the complete
galaxy, but only for the inner 11.6 kpc in radius. So the edge cutoff effect is very small
and affects the part of the potential we are looking at even less.

For the dark matter component we chose an isothermal halo with a core because of its flexibility in
fitting rotation curves. The functional form of the dark halo profile has only a second order
effect on the results of the simulations compared to the variations due to its two basic parameters
$R_c$ and $v_{\infty}$. We decided not to distinguish between different
dark halo profiles for the present analysis.

The most complex step surely is the hydrodynamical simulation of the gas flow in
the galaxy's potential. Beyond the tests of the code discussed in Section \ref{hydrosim}, it is
the excellent morphological agreement between the simulated gas density profiles and the observed
spiral arms that gives credence to the results of the code.
 The choice of the grid size was mainly motivated by the desire to
achieve reasonable computing times, and not to exceed the seeing resolution. To check the grid
cell size's effect on the results, for selected cases we also ran simulations on grids with two
times higher resolution as well as on grids with two times lower resolution. These simulations
showed some minor differences. Some gas
shocks had a slightly larger amplitude in the density distribution but their locations were
essentially unchanged. The morphology and the velocity field of the central bar was most
susceptible to the grid cell size. It results in small changes in the bar position angle.
However, we anyway exclude the central region containing the bar from our analysis.
The final $\chi^2-$fit deviation between simulations with different grid cells ranges at
about 6\% and does not change the conclusions of this paper. Accordingly we consider it safe
to perform the simulations on the medium resolution grid we used.  

In a real galaxy the assumption of a global constant pattern rotation speed may not be
fulfilled. In particular, we should expect the central bar to have a different pattern speed.
Also, the pattern might be winding slowly, rather than being fixed in a corotating frame.
Furthermore, the gas may not have a uniform temperature.
If these simplifications were relaxed, the location of the spiral arms in the
hydro-simulations would change, and eventually lead to a different overall fit quality.
Lacking any solid basis to constrain these parameters, we are unable to implement these effects
into the modelling procedure. Finally, the code does not include
gas self-gravity. The effect of gas self-gravity is difficult to quantify without actually
performing simulations, but from the literature we know that it tends to amplify the gas response.
Gas self-gravity also suppresses the tendency of the gas to shock \citep{lub86}. Since we are 
interested in the strength of the gas response to the gravitational potential and we already
find that for high disk M/L the response is too strong, we assume that our upper limit holds
also if gas self-gravity was included.  
Given the good morphological match, we may confidently assume that all
our approximations are not far off.

This leaves non-gravitationally induced gas motions as the main contamination in the comparison
of the models to the observed kinematics. The influence of non-gravitational gas motion is in our
case increased by the circumstance that we observe NGC 4254 from a perspective which is not far
from face-on, so any motion along the z-direction is expected to appear clearly in the spectra. 
Eliminating these wiggles from the rotation curve is not possible since we have no means
to reliably identify them. Any method we apply for excluding parts of the data will be
affected by some kind of bias.

\subsection{Is the Galaxy Suited for this Analysis?} \label{discsuit}
Finally we must consider the possibility that the galaxy we picked for our analysis might not be
as suited as it appeared to be. NGC 4254 is not the prototype of a classical grand design spiral
galaxy in optical wavelengths. However, as it can be seen in the central frame of Fig.~\ref{simgasdens},
the galaxy exhibits in the M/L corrected K'-band image mainly an $m = 2$ spiral pattern with a strong
symmetric part, that ends at $\approx 5.5\,$kpc and fainter outer extensions.
With its large angular size
and moderate inclination, NGC 4254 seems to be one of the most promising candidates for this 
kind of study in our sample. However, as discussed earlier in Section \ref{morphology}, there
are indications that this galaxy might not be as isolated and undisturbed as one might expect.
In fact, the morphology itself implies some perturbative event in its recent evolution
history: NGC 4254 shows a clear $m = 1$ mode and a lopsided disk. In this respect it was 
argued earlier that in-falling \ion{H}{1} gas clumps, which are visible in radio data and do not emit in
H$\alpha$, might be responsible
for triggering deviations from pure grand design structure \citep{pho93}. So have we reason
to believe that NGC 4254 is far from equilibrium? This is hard to tell, because on
the other hand we find plenty of arguments that a stable propagating density wave in NGC 4254 is
responsible for its morphology. This galaxy shows many similarities to the spiral galaxy NGC 5247
whose morphological and dynamical properties were discussed by \citet{pat97},
based on SPH simulations.  
We note that in our best model, the strong bisymmetric part of the K'
spiral, ends well inside the corotation radius, although fainter extensions reach out to it.
This picture is in agreement with the 4:1 SPH models of \citet{pat97}. Assuming
corotation close to the characteristic bifurcation of the arms at $\approx 5\,$kpc on the other
hand, we do not obtain satisfactory results (upper left frame in Fig.~\ref{simgasdens}).
Based on this is seems appropriate to conclude that NGC 4254 is at least close to an equilibrium
state and suited for a case study. 

\section{CONCLUSIONS}
We performed hydrodynamical simulations to predict the gas velocity field in a variety of potentials
for the spiral galaxy NGC 4254 and compared them to observations. These potentials consisted
of different combinations of luminous (non-axisymmetric spiral) and dark matter (axisymmetric)
components. The resulting gas spiral morphology reflects very accurately the morphology of the
galaxy and allows us to specify the corotation radius or the pattern speed of the spiral
structure quite precisely. It is noteworthy that within the error range given, the best matching
pattern speed does not depend on the mass fraction of the stellar disk relative to the dark halo.
For NGC 4254 we find that
the corotation lies at $7.5 \pm 1.1$ kpc, or at about 2.1 exponential K'-band disk scale lengths.
From the kinematics of the gas simulations we could rule out a maximal disk solution for NGC 4254.
Within the half-light radius the dark matter halo still has a non-negligible
influence on the dynamics of NGC 4254: specifically, our fraction f$_{\rm d} \la 0.85$ implies that $\ga 1/3$ of
the total mass within 2.2 K-band disk scale lengths is dark.
However, the comparison of the simulated gas velocity field to the observed rotation curves turned out
to be a delicate matter. The observed rotation curves show a significant number of bumps and wiggles,
presumably resulting from non-gravitational gas effects, that complicate the identification of wiggles
induced by the massive spiral arms. Therefore, beyond concluding that the disk is
less than 85\% of maximal, we were unable to specify a particular value for the disk mass
or to test the results from \citet{bot97} or \citet{CR99}. But already with
this statement we differ from the conclusions of \citet{deb00} and \citet{wei01b}, who argue that their
conclusions for maximal disks of barred galaxies also hold for non-barred spirals. Since
we only analyzed one galaxy so far it is inappropriate to state here that the centers of 
unbarred spirals, what we still consider NGC 4254 to be despite its small bar-like structure in
the very center, are generally governed by dark matter.

In the near future we will extend this analysis to more galaxies from our sample. This will
put us into the position to decide, whether single galaxies differ very much in their dark
matter content, or if the bulk of the spirals show similar characteristics.

\acknowledgements
We thank Panos Patsis and Julien Devriendt for their valuable comments and helpful conversations. 
We also thank the anonymous referee for a careful reading of the manuscript and comments
which improved the paper. TK likes to thank Greg Rudnick, Nicolas Cretton and Marc Sarzi for
sharing tips, tricks and thoughts.


\begin{thebibliography}{}

\bibitem[Aalto et al.(1999)]{aal99}
 Aalto, S., H\"uttemeister, S., Scoville, N. Z., Thaddeus, P. 1999, ApJ, 522, 165

\bibitem[Adler and Westpfahl(1996)]{adl96}
Adler, D. S.; Westpfahl, D. J. 1996, AJ, 111, 735

\bibitem[for a review cf., Athanassoula(1984)]{ath84}
 Athanassoula, E. 1984, in Physics Reports (Review Section of Physics Letters) 114,
 Nos. 5 \& 6, (Amsterdam: North-Holland Phys. Publ.), 319

\bibitem[Athanassoula et al.(1987)]{ath87}
 Athanassoula, E., Bosma, A., \& Papaioannou, S. 1987, A\&A, 179, 23

\bibitem[Beauvais and Bothun(1999)]{bea99}
 Beauvais, C., \& Bothun, G. 1999, ApJS, 125, 99

\bibitem[Bell and de Jong(2001)]{bel01}
 Bell, E. F., \& de Jong, R. S. 2001, ApJ, 550, 212

\bibitem[Bizenberger et al.(1998)]{biz98}
 Bizenberger, P., McCaughrean, M., Birk, C.,
 Thompson, D.,  \& Storz, C. 1998, SPIE, 3354, 825

\bibitem[Blumenthal et al.(1986)]{blu86}
 Blumenthal, G. R., Faber, S. M., Flores, R., \& Primack, J. R. 1986, ApJ, 301, 27

\bibitem[Bottema(1997)]{bot97}
 Bottema, R. 1997, A\&A, 328, 517

\bibitem[Broeils and Courteau(1997)]{bro97}
 Broeils, A. H., \& Courteau, S. 1997, in ASP Conf. Ser. 117,
 Dark and Visible Matter in Galaxies and Cosmological Implications,
 ed. M. Persic, \& P. Salucci, (San Francisco: ASP), 74

\bibitem[e.g., Canzian et al.(1993)]{can93}
 Canzian, B., Allen, R. J., \& Tilanus, R. P. J. 1993, ApJ, 406, 457

\bibitem[Canzian and Allen(1997)]{can97}
 Canzian, B., \& Allen, R. J. 1997, ApJ, 479, 723

\bibitem[Courteau and Rix(1999)]{CR99}
 Courteau S., \& Rix H.-W. 1999, ApJ, 513, 561

\bibitem[Debattista and Sellwood(1998)]{deb98}
 Debattista V. P., \& Sellwood J. A. 1998, ApJ, 493, L5

\bibitem[Debattista and Sellwood(2000)]{deb00}
 Debattista V. P., \& Sellwood J. A. 2000, ApJ, 543, 704

\bibitem[Dehnen and Binney(1998)]{deh98}
 Dehnen, W., \& Binney, J. 1998, MNRAS, 294, 429

\bibitem[Donner and Thomasson(1994)]{don94}
 Donner, K. J., \& Thomasson, M. 1994, A\&A, 290, 785

\bibitem[Elmegreen et al.(1992)]{elm92}
 Elmegreen, B. G., Elmegreen, D. M., \& Montenegro, L. 1992, ApJS, 79, 37

\bibitem[Englmaier and Gerhard(1997)]{eng97}
 Englmaier, P., \& Gerhard, O. 1997, MNRAS, 287, 57

\bibitem[Englmaier and Gerhard(1999)]{eng99}
 Englmaier, P., \& Gerhard, O. 1999, MNRAS, 304, 512

\bibitem[Federspiel et al.(1998)]{fed98}
 Federspiel, M., Tammann, G. A., \& Sandage, A. 1998, ApJ, 495, 115

\bibitem[Frei et al.(1996)]{fre96}
 Frei, Z., Guhathakurta, P., Gunn, J. E., \& Tyson, J. A. 1996, AJ, 111, 174

\bibitem[Fukushige and Makino(1997)]{fuk97}
 Fukushige, T., \& Makino, J. 1997, ApJ, 477, L9

\bibitem[Gonz\'alez and Graham(1996)]{gon96}
 Gonz\'alez, R. A., \& Graham, J. R. 1996, ApJ, 460, 651

\bibitem[Gnedin et al.(1995)]{gne95}
 Gnedin, O. Y., Goodman, J., \& Frei Z. 1995, AJ, 110, 1105

\bibitem[Grosb\o l et al.(2000)]{gro00}
 Grosb\o l, P. J., Block, D. L., \& Patsis, P. A. 2000, in ASP Conf.~Ser. 197,
 Dynamics of Galaxies, ed. F. Combes, G. A. Mamon, \& V. Charmandaris,
 (San Francisco: ASP), 191

\bibitem[e.g.~Kauffmann et al.(1999)]{kau99}
 Kauffmann G., Colberg J., Diaferio A., \& White S. D. M. 1999, MNRAS, 303, 188

\bibitem[Kent(1986)]{ken86}
 Kent, S. M. 1986, AJ, 91, 1301

\bibitem[Knapen and Beckman(1996)]{kna96}
 Knapen, J. H., \& Beckman, J. E. 1996, MNRAS, 283, 251

\bibitem[Kuijken(1995)]{kui95}
 Kuijken, K. 1995, in IAU Symp. Proc. 164, Stellar Populations,
 ed. P. C. van der Kruit, \& G. Gilmore, (Dordrecht: Kluwer), 195

\bibitem[Lin and Shu(1964)]{lin64}
 Lin, C. C., \& Shu, F. H. 1964, ApJ, 140, 646

\bibitem[Lubow et al.(1986)]{lub86}
 Lubow, S. H., Balbus, S. A., \& Cowie, L. L. 1986, ApJ, 309, 496

\bibitem[Maller et al.(2000)]{mal00}
 Maller, A., Simard, L., Guhathakurta, P., et al. 2000, ApJ, 533, 194

\bibitem[Miralda-Escude(2000)]{mir00}
 Miralda-Escude, J. 2000, ApJ, submitted, (astro-ph/0002050)

\bibitem[Moore(1994)]{moo94}
 Moore, B. 1994, Nature, 370, 629

\bibitem[Moore et al.(1999)]{moo99}
 Moore, B., Ghigna, S., Governato, F., et al. 1999, ApJ, 524, L19

\bibitem[Navarro et al.(1996)]{NFW1}
 Navarro J., Frenk C., \& White S. D. 1996, ApJ, 462, 563

\bibitem[Navarro et al.(1997)]{NFW2}
 Navarro J., Frenk C., \& White S. D. 1997, ApJ, 490, 493

\bibitem[Ostriker(2000)]{ost00}
 Ostriker, J. P. 2000, Phys. Rev. Lett., 84, 5258

\bibitem[Patsis et al.(1991)]{pat91}
 Patsis, P. A., Contopoulos, G., \& Grosb\o l, P. J. 1991, A\&A, 243, 373

\bibitem[Patsis et al.(1997)]{pat97}
 Patsis, P. A., Grosb\o l, P. J., \&  Hiotelis, N. 1997, A\&A, 323, 762

\bibitem[Patsis and Kaufmann(1999)]{pat99}
 Patsis, P. A., \& Kaufmann, D. E. 1999, A\&A, 352, 469
 
\bibitem[Phookun et al.(1993)]{pho93}
 Phookun, B., Vogel, S. N., \& Mundy, L. G. 1993, ApJ, 418. 113

\bibitem[Pierce and Tully(1988)]{pie88}
 Pierce, M. J., \& Tully, R. B. 1988, ApJ, 330, 579

\bibitem[Prendergast and Xu(1993)]{pre93}
 Prendergast, K. H., \& Xu, K. 1993, J. Comput. Phys., 109, 53

\bibitem[Quillen and Pickering(1997)]{qui97}
 Quillen, A. C., \& Pickering, T. E. 1997, AJ, 113, 2075

\bibitem[Quillen(1999)]{qui99}
 Quillen, A. C. 1999, in ASP Conf. Ser. 182, Galaxy Dynamics, ed. by D. Merritt, M. Valluri, \& J. Sellwood,
 (San Francisco: ASP), 251

\bibitem[Regan et al.(1996)]{reg96}
 Regan, M. W., Teuben, P. J., Vogel, S. N., \& van der Hulst, T. 1996, AJ, 112, 2549

\bibitem[Rix and Rieke(1993)]{RR93}
 Rix, H.-W., \& Rieke, M. 1993, ApJ, 418, 123

\bibitem[Rix and Zaritsky(1995)]{rix95}
 Rix, H.-W., \& Zaritsky, D. 1995, ApJ, 447

\bibitem[e.g. Roberts(1969)]{rob69}
 Roberts, W. W. 1969, ApJ, 158, 123

\bibitem[Sackett(1997)]{sac97}
 Sackett, P. 1997, ApJ, 483, 103

\bibitem[Sakamoto et al.(1999)]{sak99}
 Sakamoto, K., Okumura, S. K., Ishizuki, S., \&  Scoville, N. Z. 1999, ApJS, 124, 403

\bibitem[Salucci and Persic(1999)]{sal99}
 Salucci, P., \& Persic, M. 1999, A\&A, 351, 442

\bibitem[Schweizer(1976)]{sch76}
 Schweizer, F. 1976, ApJS, 31, 313

\bibitem[Slyz and Prendergast(1999)]{sly99}
 Slyz, A., \& Prendergast, K. H. 1999, A\&AS, 139, 199

\bibitem[Sandage and Tammann(1976)]{san76}
 Sandage A., \& Tammann, G. A. 1976, ApJ, 210, 7

\bibitem[Spergel and Steinhardt(2000)]{spe00}
 Spergel, D. N., \& Steinhardt, P. J. 2000, Phys. Rev. Lett., 84, 3760

\bibitem[Steinmetz and M\"uller(1995)]{ste95}
 Steinmetz, M., \& M\"uller, E. 1995, MNRAS, 276, 549

\bibitem[Thuan and Gunn(1976)]{thu76}
 Thuan, T. X., \& Gunn, J. E. 1976, PASP, 88, 543

\bibitem[van Albada et al.(1985)]{val85}
 van Albada, T. S., Bahcall, J. N., Begeman, K., \& Sancisi, R. 1985, ApJ, 295, 305

\bibitem[van Albada et al.(1986)]{val86}
 van Albada, T. S., \& Sancisi, R. 1986, 
 Phil. Trans. of the Roy. Soc. (London), Ser. A 320, no. 1556, 447

\bibitem[Visser(1980)]{vis80}
 Visser, H. C. D. 1980, A\&A, 88, 159

\bibitem[Weiner et al.(2001a)]{wei01a}
 Weiner, B. J., Williams, T. B., van Gorkom, J. H., \& Sellwood, J. A. 2001, ApJ, 546, 916

\bibitem[Weiner et al.(2001b)]{wei01b}
 Weiner, B. J., Sellwood, J. A., \& Williams, T. B. 2001, ApJ, 546, 931

\bibitem[for a review, see Xu(1998)]{xu98}
 Xu, K. 1998, Gas-Kinetic Schemes for Unsteady Compressible Flow Simulations, VKI report 1998-03
 von Karmann Institute Lecture Series
\end{thebibliography}
\end{document}